# HIGH TEMPORAL VARIABILITY IN THE OCCURRENCE OF CONSUMER-RESOURCE INTERACTIONS IN ECOLOGICAL NETWORKS

Daniela N. Lopez[1,2*], Patricio A. Camus[3,4], Nelson Valdivia[5,6] and Sergio A. Estay[1,7**]

[1] Instituto de Ciencias Ambientales y Evolutivas, Facultad de Ciencias, Universidad Austral de Chile. Valdivia, Chile.

[2] Programa de Doctorado en Ciencias mención Ecología y Evolución, Facultad de Ciencias, Universidad Austral de Chile, Valdivia.

[3] Departamento de Ecología Costera, Facultad de Ciencias, Universidad Católica de la Santísima Concepción, Concepción, Chile.

[4] Centro de Investigación en Biodiversidad y Ambientes Sustentables, Universidad Católica de la Santísima Concepción, Concepción, Chile.

[5] Instituto de Ciencias Marinas y Limnológicas, Facultad de Ciencias, Universidad Austral de Chile, Valdivia, Chile.

[6] Centro FONDAP de Investigación en Dinámica de Ecosistemas Marinos de Altas Latitudes (IDEAL), Universidad Austral de Chile, Valdivia, Chile

[7] Center for Applied Ecology and Sustainability, Pontificia Universidad Católica de Chile. Santiago, Chile.

* danyelalopez@gmail.com

** sergio.estay@uach.cl

**ABSTRACT**

Ecological networks are theoretical abstractions that represent ecological communities. These networks are usually defined as static entities, in which the occurrence of a particular interaction between species is considered fixed despite the intrinsic dynamics of ecological systems. Trophic interactions, in particular, are often temporally dynamic in nature, ranging from so-called first-order dynamics to cyclical. However, empirical analysis of the temporal variation of trophic interactions is constrained by the lack of data with high spatial, temporal, and taxonomic resolution. Here, we evaluate the spatiotemporal variability of multiple consumer-resource interactions of large marine intertidal rocky-shore networks. In order to generate the networks, more than 1,000 km of the coast of northern Chile was monitored seasonally over 3 years. The trophic interactions of all of the analyzed networks had low temporal persistence, which was well described by a common exponential decay in the rank-frequency relationship of consumer-resource interactions. This common pattern of low temporal persistence was evident despite the dissimilarities of environmental conditions among sites. Between-site rank correlations of frequency of occurrence of interactions ranged from 0.59 to 0.73. After removing the interactions with <50% frequency, the between-site correlations decreased to values between 0.60 and 0.28, indicating that low-frequency interactions accounted for the apparent similarities between sites. Our results demonstrate, therefore, that the communities studied were characterized by few persistent interactions and a large number of transient trophic interactions. We suggest that consumer-resource temporal asynchrony in addition to varying local environmental conditions and opportunistic foraging could be among the mechanisms generating the observed rank-frequency relationship of trophic interactions. Therefore, our results question the analysis of ecological communities as static and persistent natural entities and stress the need for strengthening the analysis of temporal variability in ecological networks and long-term studies.

# INTRODUCTION

Ecological networks have been useful theoretical tools for understanding the structural properties of ecological communities [1]. Currently, there are many studies analyzing the topological and statistical properties of ecological networks in order to link these network features to biological characteristics such as species richness, functional diversity, or productivity [2]. However, most ecological networks, and therefore biological communities, are recognized and analyzed as static or quasi-static structures of interacting species where all observed species and links are present simultaneously and where the variability of interactions is not considered. However, ecological systems have proven to be intrinsically dynamic in nature [3] [4], in which the presence and strength of species interactions vary in space and time [5] [6] [7].

Consumer-resource interactions (C-R) are one of the most important forces structuring ecological communities. The strength of C-R interactions tends to oscillate over time and space, influencing the structure and organization of the community and in some cases generating complex dynamics [8]. The ecological mechanisms that promote this variability include, but are not limited to, predator functional responses, interference competition, prey switching, variation in prey abundance, variation in productivity, and fluctuating environmental conditions [9] [10] [11] [12] [13]. Thus, ecological networks sampled at a specific time period or inferred from spatiotemporally heterogeneous data will not accurately reflect the long-term structure of the ecological community [5] [14] [15]. Until now, the long-term dynamics of C-R interactions have not been completely understood.

Population-dynamic theory predicts that, a close-up look of each population inside an ecological community environmental variation can strengthen or weaken existing internal feedback processes and so boost or dampen the oscillations of resource consumption [16] [17] [18]. Now zooming-out to the community, these changes in resource exploitation generate dynamic scenarios in which the strength of the interaction between multiple prey species is redistributed, generating a high frequency of weak links that stabilize the ecological networks [17] [19] [20]. In the same vein, food web theory suggests that diversity enhances stability only if a high frequency of weak C-R interactions dampens the destabilizing effects of strong interactions on community structure [21] [22] [23].

Also, few studies have analyzed the effect of spatial variation or gradients on the structure of complex network [24] [25] [26]. This variation can be represented by differences in the quality and availability of resources, productivity, consumer behavior and demography impacts consumer-resource dynamics [27]. Recent research has confirmed that trophic interactions within communities constrain the trophic structure more strongly and consistently than other factors such as dispersion [28].

Until now it has been difficult to empirically evaluate the variability and persistence of ecological interactions in real networks due to the lack of data with high spatial, temporal, and taxonomic resolution. This in turn has hampered the testing of ecological hypotheses concerning the relative frequency of occurrence of C-R interactions. Here, we analyze the spatiotemporal frequency of occurrence of C-R interactions of large intertidal rocky-shore communities sampled at a high spatial, temporal, and taxonomic resolution. Assuming that a community can be characterized by a high frequency of weak interactions or a low frequency of strong interactions, we predict that most of the C-R interactions in this model system will show low temporal persistence; that is, most interactions will occur few times over the course of the study. Our results provide new insights on the nature, structure, and dynamics of ecological communities.

**MATERIALS AND METHODS**

*Biological and environmental oceanographic data*

We analyzed a long-term dataset of consumer-resource interactions of macrobenthic (i.e. > 5 cm) intertidal rocky-shore communities in northern Chile. Four sites were sampled seasonally every three months between austral spring 2004 and summer 2007 (10 sampling times in total) along ca.1,000 km of the northern coast of Chile: Río Seco (RS, 21.00°S, 70.17°W), Caleta Constitución (CC, 23.42°S, 70.59°W), Caleta Angosta (CA, 28.26°S, 71.38°W) and Lagunillas (LA, 30.10°S, 71.38°W).

In general, across the four sites we identified over 200 prey species, 41 consumers, and more than 1,000 interactions. See sampling and methodology details in [29] [30]. Environmental conditions of each site were assessed through publicly available sea surface satellite data: mean values of sea surface temperature, dissolved oxygen, salinity, and chlorophyll a concentration were obtained for the pixels nearest to the coast from WorldClim Global Climate Data [31] and MARSPEC [32], with temporal resolutions of 50 and 55 years, respectively (Table 1).

In each site, at least ten individuals of each carnivorous and herbivorous species were collected considering that a sample of 10 individuals represents > 85% of the population diet [30]. After collection, samples were subsequently fixed in 10% formalin. In the laboratory, the prey items of each consumer individual were identified to the lowest possible taxonomic level by means of gut content examination [29] [30]. For each site, we constructed a consumer-resource matrix with the number of times that each C-R interaction was present in the 10 sample replicates (i.e. a "persistence" matrix). This was done using the vector sum of the ten original presence/absence matrices of each site. The final persistence matrix contained values from 1 to 10, where 1 means that the interaction was observed once (i.e. low persistence), and 10 means that the interaction was present in all sample replicates (i.e. high persistence). In addition, we determined species richness as the number of taxonomic identities (number of prey and consumers, Table 1). Finally, we computed a dissimilarity matrix (1-Jaccard index) from the presence/absence data in order to determine the degree to which diet composition varied between sampling sites [33].

*Statistical analysis*

The prediction that most of the C-R interactions in this model system would have low temporal persistence in the long term was tested by fitting a rank-ordered function to the observed distribution of persistence (ranging from 1 to 10) of the C-R interactions. As a base for parameter estimation, we used a discrete version of the generalized beta distribution (DGBD) [34] [35]. The

DGBD has the form:

$$f(r) = A(N+1-r)^b/(r)^a,$$

where r is the ranked value of persistence of the C-R interactions, N is the maximum possible value of the rank (10), A is a normalization constant, and *a* and *b* are power law exponents. The exponent *a* is related to the "left to right" tail and the exponent *b* is related to the "right to left" tail of the dataset [35] [36] [37]. Different combinations of values of *a* and *b* correspond to different shapes of the curve. If *a* < *b*, the curve shows a smooth exponential decay because the frequencies of occurrence are similar across ranks. If *a* > *b*, the curve shows a strong exponential decay because the left tail of the curve is significantly larger than the right tail.

To evaluate the degree to which each C-R interaction had similar values of persistence between communities, we computed Spearman rank correlations between the site-specific persistence matrices. All matrices contained the same C-R interactions and were sorted in the same way. Interactions that were absent in a given site scored zero for persistence. A high correlation coefficient could be expected if each interaction presented similar persistence values between sites. After the analysis, we sequentially recalculated the correlation coefficient after consecutively removing the least frequent interactions (i.e. occurring < 5 times) and the most persistent interactions (i.e. occurring ≥ 5 times).

## RESULTS

*Persistence in the occurrence of C–R interactions*

Our results showed a total of 5120 trophic interactions were detected during the study (Fig. 1). Less than 1% of the C-R interactions were consistently observed over time in all of the four communities; more than 40% of the interactions occurred only once (Fig. 1). In addition, <16% of the interactions were detected in more than 5 sample replicates, and >80% of the interactions occurred less than 5 times (Fig. 1). Accordingly, we observed a strong exponential decay in the rank-frequency relationship of interaction persistence for all sites, and a high goodness-of-fit for DGBD was found (Fig. 2). From north to south, the coefficients of the DGBD were as follows: Río Seco $a = 1.15$, $b = 0.62$, $R^2 = 0.97$; Caleta Constitución $a = 0.99$, $b = 0.92$, $R^2 = 0.99$; Caleta Angosta $a = 0.99$, $b = 1.05$, $R^2 = 0.99$ and Lagunillas $a = 1.3$, $b = 0.65$, $R^2 = 0.99$. The exponential decay of the rank-frequency distribution of interaction persistence was common for all communities (Fig. 1) despite there being relatively large environmental differences in temperature, chlorophyll a, and biotic aggregate properties such as species richness (Table 1). However, pairwise comparisons of sites in terms of species composition suggested that they were relatively similar (Jaccard (dis)similarities: 68% - 77%; Fig. S1).

*Correlation of trophic interactions between communities*

The between-site Spearman rank correlations ranged from 0.59 to 0.73, suggesting that C-R interaction persistence was similar between sites. After removing the interactions observed <5 times, however, the between-site correlations decreased to values between 0.60 and 0.28 (Fig. 3, Fig. S2). This result indicates that the large number of less persistent interactions accounted for the high between-site correlations.

**DISCUSSION**

According to these results, we observed a common decay in the rank-frequency distribution of interaction persistence for intertidal rocky-shore assemblages spanning more than 1,000 km of the Chilean coast. This common pattern was observed despite the relatively large differences in environmental conditions among sites. The parameters of the rank-frequency functions were similar across all communities; these parameters were almost always $a > b$. One exception to this included Caleta Angosta where $b > a$ indicating that different communities had similar trophic structures, and there is probably a common underlying mechanism responsible for the observed rank-frequency function's skewed distribution of interaction persistence [37]. The apparent similarity among sites was accounted for by a large number of interactions occurring in less than 50% of the sample replicates. Regarding these communities, Camus et al. [30] suggest that core prey species are consistently consumed at high frequencies in time and space. In our study, only 23 interactions—i.e. <1% of a total of 5120—occurred in all of the 10 sampling times. Therefore, the structure of these food webs is characterized by a low number of persistent interactions, and a large number of transient, low-persistence interactions. Our work is one of the few empirical studies demonstrating that food webs are dominated by low-frequency trophic interactions (also see ref. [4]).

Seasonal variation in the abundance of prey and predators might explain the low interaction persistence reported in this study. The communities of this study have been shown to have high seasonal turnover in species composition (resources and prey)—for example 40 to 50% of the species are replaced from one season to another [38]. This high turnover could generate seasonal asynchrony between consumer and prey populations, reducing their encounter probability and therefore generating the low persistence of C-R interactions here described. Nevertheless, other sources of variation such as environmental fluctuations affecting growth rates, mortality, reproduction and foraging behavior [39], in addition to seasonal fluctuations in rates of primary production [40] cannot be ruled out. Therefore, multifactorial studies should be conducted in order to assess the independent and combined effects of multiple factors that influence the persistence of trophic interactions.

We observed a high correlation in the frequency of occurrence of interactions between communities, but these correlations decreased after we removed the least persistent interactions (i.e. occurring < 5 times) from the analysis. Firstly, this result indicates that the similarity of persistence of trophic interactions was due to the large number of less persistent interactions. Secondly, the most persistent interactions had different frequencies of occurrence among sites. For example, the trophic interaction between the gastropod *Tegula atra* and encrusting Corallinaceae algae was one of the few interactions observed in all of the sample replicates, yet this interaction only occurred in three our of the four sites. The responses of consumer behavior to local environmental conditions could well explain the geographical variation in the frequency of occurrence of a given interaction. For instance, foraging behavior can be affected by nutrient concentration and temperature [41] [42]. In our study, local environmental conditions varied among sites: Caleta Constitución and Lagunillas are characterized by the persistent influence of an upwelling event, while Río Seco and Caleta Angosta are less influenced by this source of nutrient-rich and cold waters [43] [44]. In central Chile [45] and the Pacific coast of North America [46], upwelling activity has been shown to significantly affect foraging rates of intertidal limpets and snails, respectively. Moreover, regional-scale analyses demonstrate that predation rate on prey adults and recruits peaks at intermediate levels of upwelling constancy [47]. Accordingly, it is hypothesized that local environmental factors and interaction persistence are significantly interdependent . Testing this hypothesis could shed light on the community-level consequences of drastic environmental changes.

The low-frequency interactions, which represented >80% of the whole set of interactions of each community, could have a role in maintaining community stability. For example, transient interactions can result from behavioral prey shifts that occur over short time scales, which can dampen the destabilizing effects of temporal mismatches between consumers and resources [22]. If this is true, the presence of a high number of low-persistence trophic interactions might imply that the network is robust to random disturbances (i.e. random removal or addition of species), but susceptible to non-random eliminations of highly connected or constantly dominant species. Thus, we can hypothesize that the maintenance of a large number of transient interactions is pivotal for the stability of the analyzed networks. Community stability has been generally linked with interaction strength; weaker interactions increase stability [34]. To our best knowledge, however, interaction persistence has not been included into the community stability theory. Our results stress the role of the temporal variation and persistence of species interactions in maintaining the structure and stability of local communities.

In summary, our study shows a general pattern of ecological networks comprised of a low number of persistent interactions, but a large number of transient, low-persistence species interactions. Most ecological networks are analyzed as static structures and few empirical studies have dealt with the temporal and spatial variability of those networks. Our results challenge the analysis of ecological communities as static and persistent natural entities—where species and links are present simultaneously—and stress the need for the analysis of the processes that generate significant temporal variation in ecological networks. The relevance of long-term ecological data to understanding how ecological networks respond to ongoing environmental scenarios cannot be underestimated.

## ACKNOWLEDGEMENTS

We thank financial support from CONICYT grant N° 21140959 (DNL), FONDECYT N° 1040425 (PAC), 1141037 (NV) and Center of Applied Ecology & Sustainability (CAPES) FB 0002 (SAE). NV was supported by FONDAP IDEAL grant Nº 15150003.

## REFERENCES

1. Mitchell, M. 2009 *Complexity: A guided tour*. Oxford University Press.

2. Poisot, T., Stouffer, D. B. & Gravel, D. 2015 Beyond species: why ecological interaction networks vary through space and time. *Oikos* **124**, 243–251.

3. Thompson, R. M. et al. 2012 Food webs: reconciling the structure and function of biodiversity. *Trends Ecol. Evol.* **27**, 689–97. (doi:10.1016/j.tree.2012.08.005)

4. Rasmussen, C., Dupont, Y. L., Mosbacher, J. B., Trøjelsgaard, K. & Olesen, J. M. 2013 Strong impact of temporal resolution on the structure of an ecological network. *PLoS One* **8**, e81694. (doi:10.1371/journal.pone.0081694)

5. Poisot, T., Canard, E., Mouillot, D., Mouquet, N., Gravel, D. & Jordan, F. 2012 The dissimilarity of species interaction networks. *Ecol. Lett.* **15**, 1353–61. (doi:10.1111/ele.12002)

6. Winemiller, K. O. 1990 Spatial and temporal variation in tropical fish trophic networks. *Ecol. Monogr.* **60**, 331–367.

7. Schoenly, K. & Cohen, J. E. 1991 Temporal variation in food web structure: 16 empirical cases. *Ecol. Monogr.* , 267–298.

8. Pascual & Dunne 2005 *Ecological Networks: Linking Structure to Dynamics in Food Webs*.

9. Akin, S. & Winemiller, K. O. 2006 Seasonal variation in food web composition and structure in a temperate tidal estuary. *Estuaries and Coasts* **29**, 552–567.

10. Berlow, E. L. et al. 2004 Interaction strengths in food webs: issues and opportunities. *J. Anim. Ecol.* **73**, 585–598.

11. Van der Putten, W. H., de Ruiter, P. C., Bezemer, T. M., Harvey, J. A., Wassen, M. & Wolters, V. 2004 Trophic interactions in a changing world. *Basic Appl. Ecol.* **5**, 487–494.

12. Kaartinen, R. & Roslin, T. 2012 High temporal consistency in quantitative food web structure in the face of extreme species turnover. *Oikos* **121**, 1771–1782.

13. Zook, A. E., Eklof, A., Jacob, U. & Allesina, S. 2011 Food webs : Ordering species according to body size yields high degree of intervality. *J. Theor. Biol.* **271**, 106–113. (doi:10.1016/j.jtbi.2010.11.045)

14. Thompson, R. M. & Townsend, C. R. 1999 The effect of seasonal variation on the community structure and food-web attributes of two streams: implications for food-web science. *Oikos* **87**, 75–88.

15. Hiltunen, T., Jones, L. E., Ellner, S. P. & Hairston Jr, N. G. 2013 Temporal dynamics of a simple community with intraguild predation: an experimental test. *Ecology* **94**, 773–779.

16. Amarasekare, P. 2014 Effects of temperature on consumer-resource interactions. *J. Anim. Ecol.* (doi:10.1111/1365-2656.12320)

17. Ings, T. C. et al. 2009 Ecological networks--beyond food webs. *J. Anim. Ecol.* **78**, 253–69. (doi:10.1111/j.1365-2656.2008.01460.x)

18. Gilbert, A. J. 2009 Connectance indicates the robustness of food webs when subjected to species loss. *Ecol. Indic.* **9**, 72–80. (doi:10.1016/j.ecolind.2008.01.010)

19. Kondoh, M. 2003 Foraging adaptation and the relationship between food-web complexity and stability. *Science* **299**, 1388–1391.

20. Loeuille, N. 2010 Influence of evolution on the stability of ecological communities. *Ecol. Lett.* **13**, 1536–1545.

21. Bellmore, J. R., Baxter, C. V & Connolly, P. J. 2015 Spatial complexity reduces interaction strengths in the meta-food web of a river floodplain mosaic. *Ecology* **96**, 274–283.

22. McCann, K., Hastings, A. & Huxel, G. R. 1998 Weak trophic interactions and the balance of nature. *Nature* **395**, 794–798.

23. McCann, K. S. 2000 The diversity–stability debate. *Nature* **405**, 228–233. (doi:10.1038/35012234)

24. Martinez, N. D. 1991 Artifacts or attributes? Effects of resolution on the Little Rock Lake food web. *Ecol. Monogr.* **61**, 367–392.

25. Closs, G. P. & Lake, P. S. 1994 Spatial and temporal variation in the structure of an intermittent-stream food web. *Ecol. Monogr.* **64**, 2–21.

26. Woodward, G. et al. 2010 Ecological Networks. **42**. (doi:10.1016/B978-0-12-381363-3.00002-2)

27. Polis, G. A. & Strong, D. R. 1996 Food web complexity and community dynamics. *Am. Nat.* **147**, 813–846.

28. Morlon, H., Kefi, S. & Martinez, N. D. 2014 Effects of trophic similarity on community composition. *Ecol. Lett.* **17**, 1495–1506.

29. Camus, P. & Daroch, K. 2008 Potential for omnivory and apparent intraguild predation in rocky intertidal herbivore assemblages from northern Chile. *Mar. Ecol. Prog. Ser.* **361**, 35–45.

30. Camus, P. A., Arancibia, P. A. & Ávila-Thieme, I. 2013 A trophic characterization of intertidal consumers on Chilean rocky shores. *Rev. Biol. Mar. Oceanogr.* **48**, 431–450. (doi:10.4067/S0718-19572013000300003)

31. Hijmans, R. J., Cameron, S. E., Parra, J. L., Jones, P. G. & Jarvis, A. 2005 Very high resolution interpolated climate surfaces for global land areas. *Int. J. Climatol.* **25**, 1965–1978.

32. Sbrocco, E. J. & Barber, P. H. 2013 MARSPEC: ocean climate layers for marine spatial ecology: Ecological Archives E094-086. *Ecology* **94**, 979.

33. Dornelas, M. et al. 2013 Quantifying temporal change in biodiversity: challenges and opportunities. *Proc. Biol. Sci.* **280**, 20121931. (doi:10.1098/rspb.2012.1931)

34. Zhang, J. & Feng, Y. 2014 Common patterns of energy flow and biomass distribution on weighted food webs. *Phys. A Stat. Mech. its Appl.* **405**, 278–288. (doi:http://dx.doi.org/10.1016/j.physa.2014.03.040)

35. Martínez-Mekler, G., Martínez, R. A., Del Rio, M. B., Mansilla, R., Miramontes, P. & Cocho, G. 2009 Universality of rank-ordering distributions in the arts and sciences. *PLoS One* **4**, e4791–e4791.

36. Sornette, D. 2006 *Critical Phenomena in Natural Sciences: Chaos, Fractals, Selforganization and Disorder: Concepts and Tools*. Second Edi. Springer Series in Synergetics.

37. Finley, B. J. & Kilkki, K. 2014 Exploring Empirical Rank-Frequency Distributions Longitudinally through a Simple Stochastic Process. *PLoS One* **9**, e94920.

38. Camus, P. A. 2008 Understanding biological impacts of ENSO on the eastern Pacific: An evolving scenario. *Int. J. Environ. Health.* **2**, 5–19.

39. Crone, E. M. & E. 2013 The role of transient dynamics in stochastic population growth for nine perennial plants.

40. Menge, B. A. 2000 Top-down and bottom-up community regulation in marine rocky intertidal habitats. *J. Exp. Mar. Bio. Ecol.* **250**, 257–289.

41. Pool, T. K., Cucherousset, J., Boulêtreau, S., Villéger, S., Strecker, A. L. & Grenouillet, G. 2016 Increased taxonomic and functional similarity does not increase the trocommunities and most occurred oncephic similarity of communities. *Glob. Ecol. Biogeogr.* **25**, 46–54.

42. Moore, J. C. & de Ruiter, P. C. 1991 Temporal and spatial heterogeneity of trophic interactions within below-ground food webs. *Agric. Ecosyst. Environ.* **34**, 371–397.

43. Camus, P. A. & Andrade, Y. 1999 Diversidad de comunidades intermareales rocosas del norte de Chile y el efecto potencial de la surgencia costera. *Rev. Chil. Hist. Nat.* **72**, 389–410.

44. Thiel, M. et al. 2007 The Humboldt Current System of northern and central Chile: oceanographic processes, ecological interactions and socioeconomic feedback. *Oceanogr. Mar. Biol. Ann. Rev.* **45**, 195–344.

45. Nielsen, K. J. & Navarrete, S. A. 2004 Mesoscale regulation comes from the bottom-up: intertidal interactions between consumers and upwelling. *Ecol. Lett.* **7**, 31–41.

46. Sanford, E., Roth, M. S., Johns, G. C., Wares, J. P. & Somero, G. N. 2003 Local selection and latitudinal variation in a marine predator-prey interaction. *Science* **300**, 1135–1137.

47. Menge, B. A. & Menge, D. N. L. 2013 Dynamics of coastal meta-ecosystems: the intermittent upwelling hypothesis and a test in rocky intertidal regions. *Ecol. Monogr.* **83**, 283–310.

**Figure caption**

**Figure 1:** Frequency of occurrence of consumer-resource (C-R) interactions. The columns represent consumers and rows represent resources. Each cell represents a trophic interaction. The color scale indicates less to more persistent interactions (white and red, respectively). Total number of sample replicates was 10.

**Figure 2:** Rank-ordered distributions of the persistence of trophic interactions in four northern Chilean sites: Río Seco (RS, grey), Caleta Constitución (CC, green), Caleta Angosta (CA, blue) and Lagunillas (LA, red). Solid lines are DGDB fits with ($a$, $b$) = Río Seco $a$ = 1.15, $b$ = 0.62, Caleta Constitución $a$ = 0.99, $b$ = 0.92, Caleta Angosta $a$ = 0.99, $b$ = 1.05 and Lagunillas $a$ = 1.3, $b$ = 0.65. The insert shows the values of parameter $a$ in the DGDB and the 95% confidence intervals (95%).

**Figure 3:** Correlograms of between-site Spearman rank correlations of C-R persistence matrices. Site codes are: Río Seco (RS), Caleta Constitución (CC), Caleta Angosta (CA) and Lagunillas (LA). The circle color scale indicates the strength of the correlation (correlation coefficient). Left panel shows the correlations between matrices that included all interactions. The right panel shows the between-site correlations after the less persistent interactions (occurring < 5 times) were removed from the dataset.

**Figure 1.**

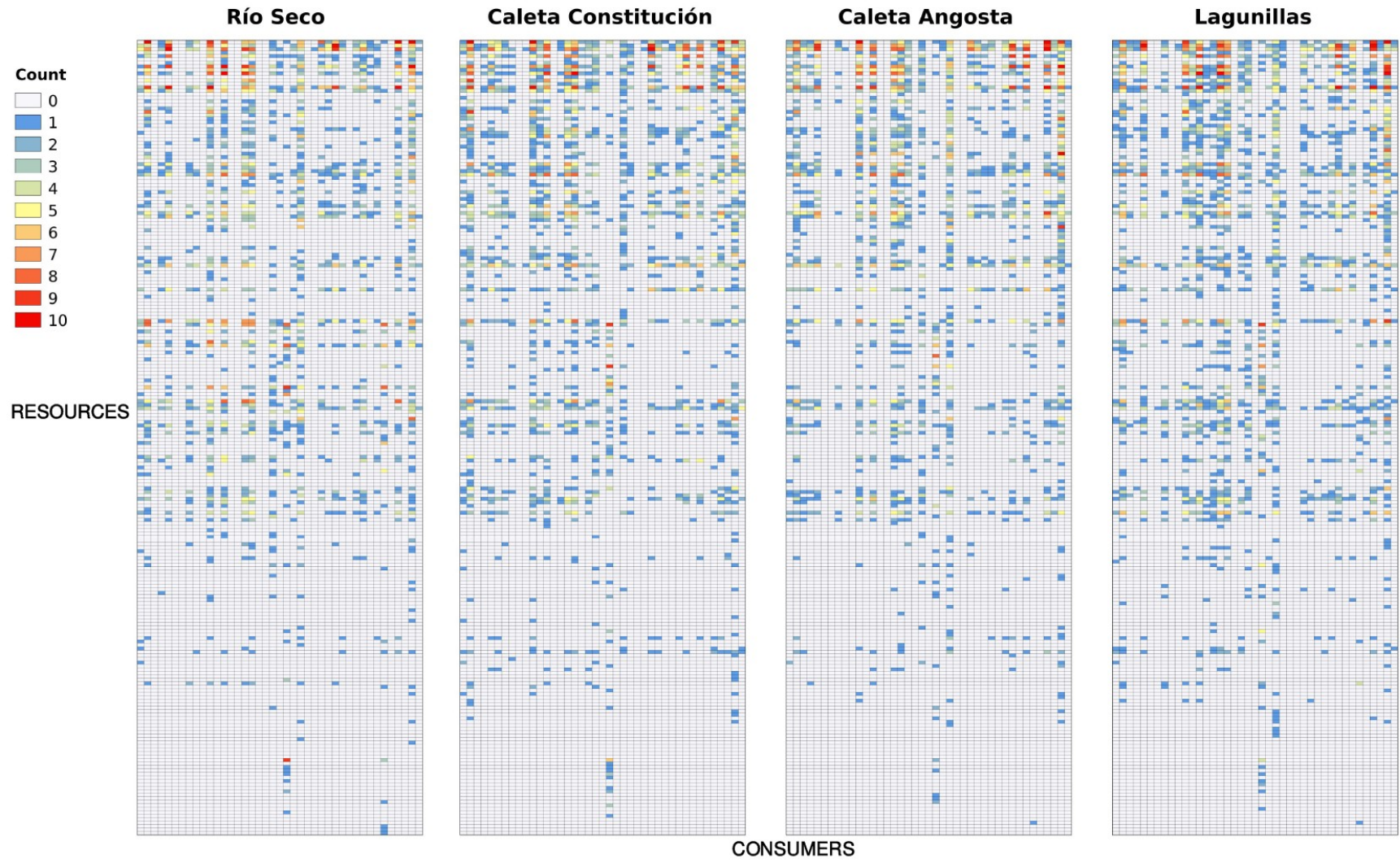

**Figure 2.**

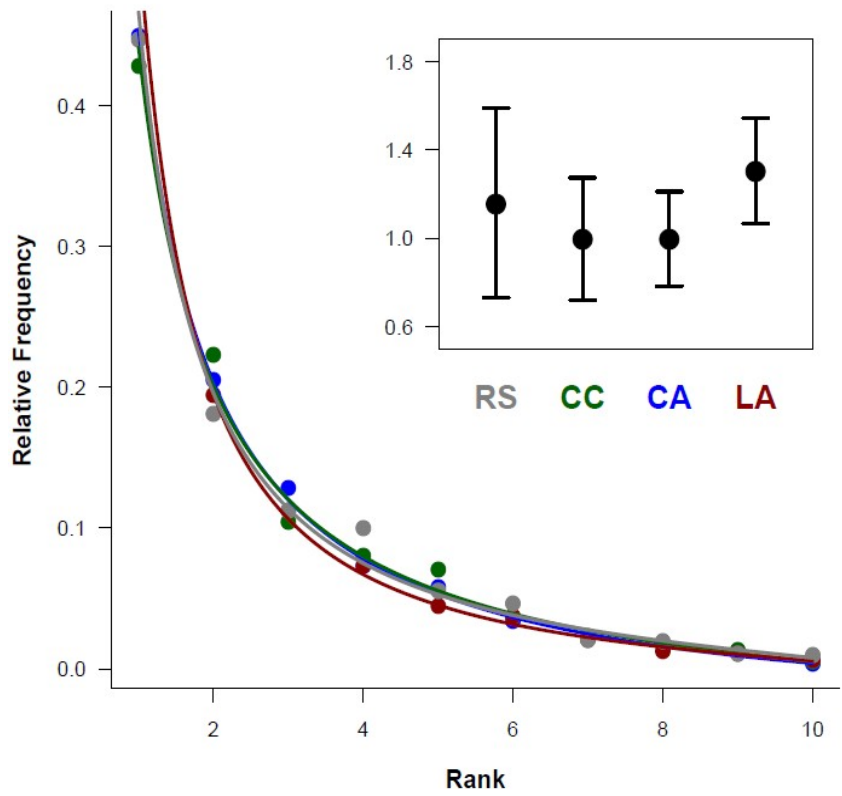

**Figure 3.**

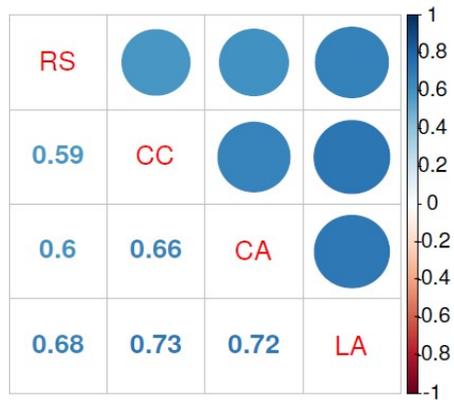 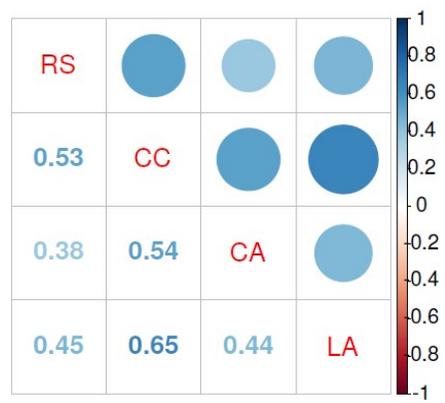

**Tables**

Table 1: Mean sea surface temperature (SSTmean), mean sea dissolved-oxygen (Dissox), mean sea chlorophyll a (CHLOmean), mean sea salinity and the number of intertidal rocky-shore species (Species Richness) for each sampling site in northern Chilean (Río Seco, Caleta Constitución, Caleta Angosta and Lagunillas). Environmental layers were extracted from WorldClim - Global Climate Data and MARSPEC (ocean climate layers for marine spatial ecology). Richness included consumer and prey species.

| Sites | SSTmean | Dissox | CHLOmean | Salinity | Species Richness |
|---|---|---|---|---|---|
| Río Seco | 18.012 | 5.192 | 6.011 | 34.898 | 168 |
| Caleta Constitución | 16.971 | 5.509 | 0.636 | 34.757 | 175 |
| Caleta Angosta | 15.683 | 5.451 | 1.755 | 34.616 | 192 |
| Lagunillas | 15.446 | 5.503 | 4.531 | 34.464 | 199 |

**Supplementary material**

**HIGH SPATIO-TEMPORAL VARIABLITY IN THE OCCURENCE OF CONSUMER-RESOURCE INTERACTIONS IN ECOLOGICAL NETWORKS**

Lopez, Daniela N[1,2], Camus, Patricio A[3], Valdivia, Nelson[4,5] & Estay, Sergio A[1,6]

[1] Instituto de Ciencias Ambientales y Evolutivas, Facultad de Ciencias, Universidad Austral de Chile. Valdivia, Chile.

[2] Programa de Doctorado en Ciencias mención Ecología y Evolución, Facultad de Ciencias, Universidad Austral de Chile, Valdivia.

[3] Departamento de Ecología Costera, Facultad de Ciencias, Universidad Católica de la Santísima Concepción, Concepción, Chile.

[4] Instituto de Ciencias Marinas y Limnológicas, Facultad de Ciencias, Universidad Austral de Chile, Valdivia, Chile.

[5] Centro de Investigación Ecosistemas Marinos de Altas Latitudes (IDEAL), Universidad Austral de Chile, Valdivia, Chile

[6] Center for Applied Ecology and Sustainability, Pontificia Universidad Católica de Chile. Santiago, Chile.

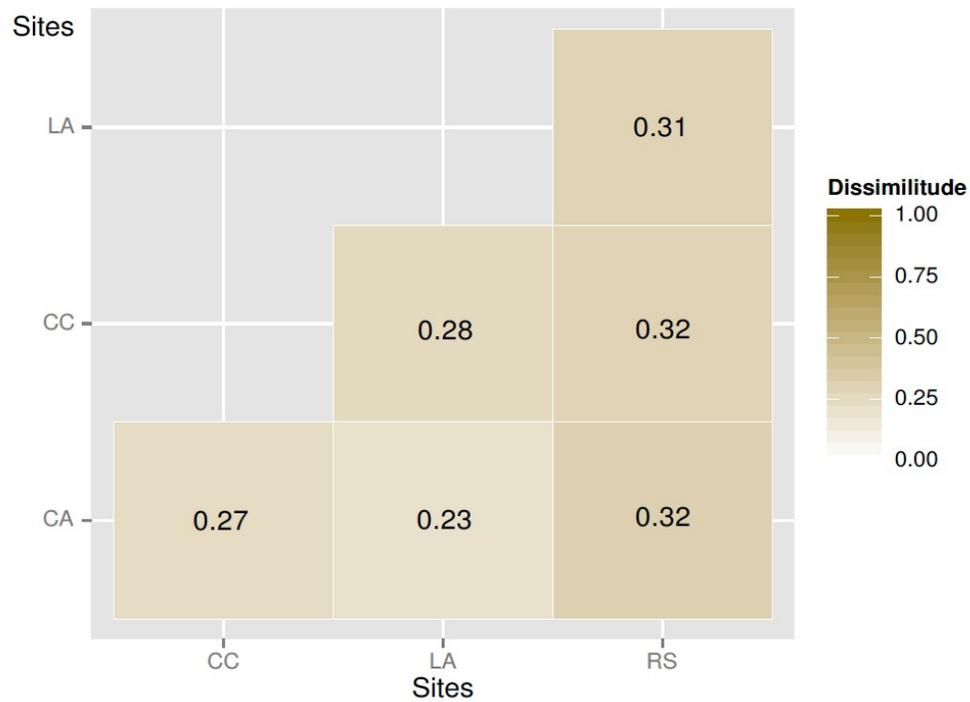

Fig.S1: Between-site Jaccard dissimilarity in species composition of diets in four northern Chilean sites. 0 indicates maximum similarity in species composition and 1 represents the opposite. Site codes are: Río Seco (RS), Caleta Constitución (CC), Caleta Angosta (CA) and Lagunillas (LA)

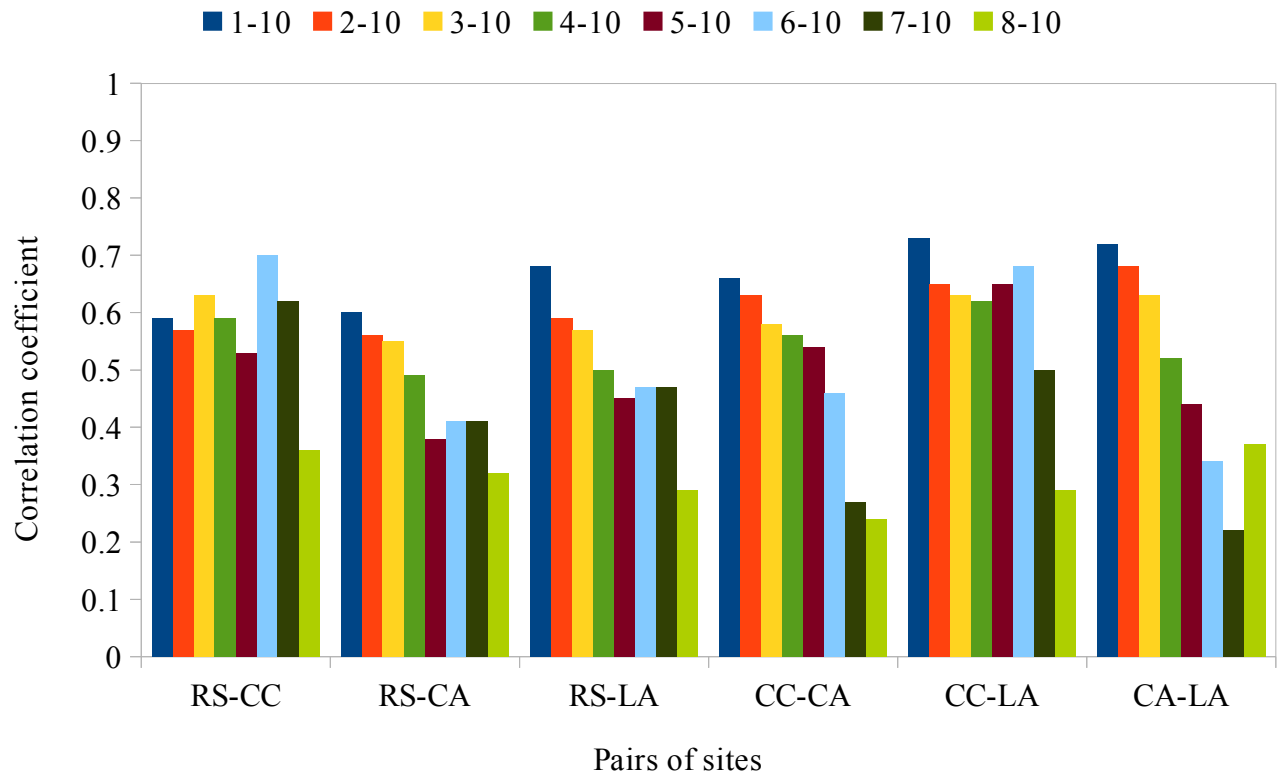

Fig. S2: Correlation coefficients of trophic interactions between pair sites after removing least frequent interactions and most persistent interactions. Site codes are: Río Seco (RS), Caleta Constitución (CC), Caleta Angosta (CA) and Lagunillas (LA). Bar height indicates the strength of the correlation and values indicate the correlation coefficient. The series 1-10 (blue) indicates that the correlation was calculated with the interactions that were present in 1 to 10 replicates; the same applies to the rest of the series.